\documentclass[twocolumn]{book}
\usepackage[dvips]{graphicx,color}
\usepackage{universe}

\makeauthorindex
\BookTitle{Proceedings of the XXIX PHYSICS IN COLLISION}

\CopyRight{\copyright 2009 by Universal Academy Press, Inc.}

\begin{document} %******************************************

%\tableofcontents
\pagenumbering{arabic}

\chapter{%
{\LARGE \sf
Study of Double Drell-Yan Process} \\
{\normalsize \bf %%%%%%%%%%%%%%******* Authors **************
Miroslav Myska$^{1,2}$ } \\
{\small \it \vspace{-.5\baselineskip}%***** Affiliations ***********
(1) Institute of Physics, Academy of Sciences of the Czech Republic,
      Na Slovance 2, CZ-182 21, Prague, Czech Republic \\
(2) Czech Technical University in Prague, FNSPE, Brehova 7, CZ-115 19, Prague, Czech Republic
}
}

%**************************
% Please note:
% One \AuthorContents{} is necessary
% for EACH CONTRIBUTION, for the contents page and
% One \AuthorIndex{} is necessary
% for EACH AUTHOR, for the index.
%**************************

%***** Item below is the data for CONTENTS.
%***** Please enter all author's name that should be initialized.
% \AuthorContents{K.\ Kawagoe and Y.\ Yamazaki}

%***** items below are the data for AUTHOR INDEX.
%***** Please enter a author's name that should be initialized.
% \AuthorIndex{Kawagoe}{K.}
% \AuthorIndex{Yamazaki}{Y.}

  \baselineskip=10pt %*******
  \parindent=10pt    %*******

\section*{Abstract} %******** Body of document starts.****************

This study focuses on the multiple parton scattering theory as a
background for the new measurement of the inclusive cross section of
the vector boson pair production at the LHC energies. The process
under study is the double Drell-Yan annihilation. In this case, two
quark-antiquark annihilations occur independently in one
proton-proton annihilation. The final state with two pairs of
leptons (electron or muon pair) is investigated while the
intermedial vector boson can be both gamma and Z.

\section{Introduction} %%%%%%%%%%%%%%%

The traditional hadron collider experiment studies are
usually based on the concept where one parton from every
hadron takes place in the head-on collision of the
colliding particles. With larger energies, the accelerated
hadron is a composite object consisting of many partons
from the sea. This increases the probability that two
or more independent hard parton-parton scatterings can
simultaneously appear in the same hadron-hadron interaction.
This mechanism is generally called multiple parton scattering (MPS).

\begin{figure}[h!]
\begin{center}
\includegraphics[width=6cm]{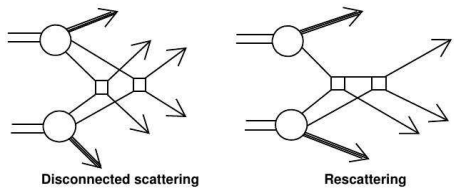}
\caption{Two kinds of multiple parton scattering: the disconnected
scattering and the rescattering.}
\label{fig:myska_rescattering}
\end{center}
\end{figure}

In fact, there are two kinds of processes called multiple
parton scattering, see Fig. \ref{fig:myska_rescattering}
The disconnected scattering was described in the previous paragraphs.
During rescattering, one parton from the first-beam hadron interacts
with two or more partons from the second-beam
hadron. This process is significant for heavy ion collisions
but is highly suppressed at proton-proton collisions
and will not be further involved.

If all the parton interactions occur independently then
the number of hard scatterings $h$ at a given impact parameter
$b$ and CMS energy $\sqrt{s}$ of the given hadron-hadron
interaction follows the Poissonian statistics:

\begin{equation}
P_h(b,s) = \frac{<n(b,s)>^h}{h!}e^{<n(b,s)>},
\end{equation}

\noindent where $<n(b,s)>$ is the average number hard scatterings
and can be later expressed as (7).

The additional parton scatterings proceed mostly via
soft or semi-hard interactions at energies that have been
reached up to now, but the chance of hard scattering will
rise with the increase of the c.m.s. energy (like 7, 10,
and 14 TeV planned for the LHC) and even faster than
the classical single parton scattering production cross
section.

\section{Double Parton Scattering} %%%%%%%%%%%%%%%

\begin{figure}[h!]
\begin{center}
\includegraphics[width=4cm]{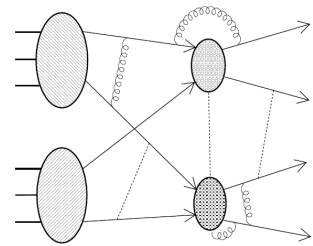}
\caption{Schematic picture of double parton scattering with possible
virtual gluon connections.}
\label{fig:myska_dps}
\end{center}
\end{figure}

The double parton scattering (DPS) is the simplest case of the MPS.
Figure \ref{fig:myska_dps} shows two unspecified hard parton
interactions taking part in one hadron-hadron interaction, and they
are not connected or correlated in any kinematic observable. The
processes are separated in transverse space by a distance of the
magnitude of the hadron radius. The source of possible correlations
would be an emission of virtual connecting (C) gluons by interacting
partons. The C-gluons connect the hard processes (dotted lines in
Fig. \ref{fig:myska_dps}), while the non-connecting (NC) gluons
contribute to each hard process separately (curly lines in Fig.
\ref{fig:myska_dps}). The CDF collaboration analyzed a 16
$pb^{−1}$ sample of $p\bar{p} \rightarrow \gamma/\pi^0 + 3jets +
X$ data \cite{CDF} and did not find any apparent evidence of
correlations in parton momentum fractions x, nor any evidence of
kinematic correlations in $mass$, $p_T$, or $p_z$.

\begin{figure}[h!]
\begin{center}
\includegraphics[width=4.5cm]{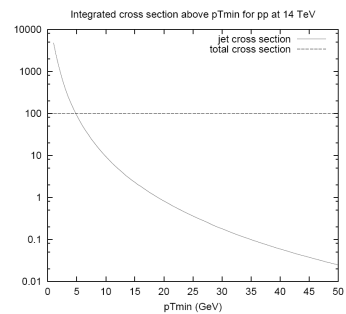}
\caption{The rise of the jet-production cross section over the total
cross section at a low transverse momentum region.}
\label{fig:myska_unitarity}
\end{center}
\end{figure}

The multiple parton scattering processes offer a solution
for the unitarity violation of the total integrated
cross section for hard collisions which is an already
known fact appearing in the region near the perturbative
threshold. The total cross section calculated from
the parton model, assuming only single parton scattering,
predicts a constant rate in the whole perturbative
transverse momentum range. However, there are many
processes, especially multi-jet production, contributing
over the predicted value, see Fig. \ref{fig:myska_unitarity}
This increased activity
can be explained by assuming the MPS.

For instance, DPS can produce the same final state
(and with a comparable production rate) as the standard
single parton process with addition of perturbative
corrections (multi-jets events) or by t-channel quark exchange
(IVB boson pair production).

The importance of the MPS lies also in its ability to
provide information on the spatial distribution of partons
within the proton, as well as all other possible correlations
among them. Its observation could bring a new
parton concept complementary to the common description,
through the single-parton distribution functions.

The DPS cross section can be factorized by the double
parton distribution function $\Gamma(x_1,x_2;b)$ as a function of
two momentum fractions of the interacting partons and
of their relative transverse distance $b$:\\

\noindent $\sigma_D = \frac{m}{2}\int_{p_T^{c}}\Gamma_A(x_1,x_2;b)\hat{\sigma}(x_1,x_1')
\hat{\sigma}(x_2,x_2')\Gamma_B(x_1',x_2';b)$

\begin{equation}
\times dx_1dx_1'dx_2dx_2',
\label{eq:myska_factorization}
\end{equation}

\noindent where $m = 1$ for indistinguishable partons processes and
$m = 2$ for distinguishable partons processes.

\section{Hadron Matter Distribution}

The above mentioned DPS cross section formula
\ref{eq:myska_factorization} contains the double parton distribution
functions. Current models, based on the inclusiveness of the
standard parton distribution functions (PDF), allow the double
parton distributions to be factorized by standard PDF's in
convolution with the function of b expressing the spatial
distribution of partons in the transverse plane and their overlap:

\begin{equation}
 \Gamma(x_1,x_2;b) = f(x_1)f(x_2)F(b).
\label{eq:myska_distrib}
\end{equation}

The next step is to express the total cross section,
where the number of parton interactions would appear
as the index of the sum from one to infinity. Using
the optical theorem and neglecting spin effects, the total
cross section can be related to the imaginary part of the
Fourier transform of the elastic scattering amplitude in
the impact parameter $a(b,s)$:

\begin{equation}
\sigma_{tot} = 4\pi \int d^2b Im \left( a(b,s) \right).
\end{equation}

The MPS model also stems from the eikonal approximation
framework. Here, the elastic amplitude can be
written down in terms of the eikonal function $\chi(b,s)$ as

\begin{equation}
 a(b,s) = \frac{e^{-\chi(b,s)}-1}{2i}
\end{equation}

\noindent and similarly the inelastic part of the total cross section:

\begin{equation}
 \sigma_{hard} = \pi \int_0^{\infty}d^2b \left[ 1-e^{-2\chi(b,s)} \right].
\end{equation}

The assumption of the independency of individual parton
interactions leads to the Poissonian model, as mentioned
above. The average number of secondary hard
scatterings is thus given by:

\begin{equation}
 <n(b,s)> = F(b)\sigma_{hard}(s).
\end{equation}

\begin{figure}[h!]
\begin{center}
\includegraphics[width=4.5cm]{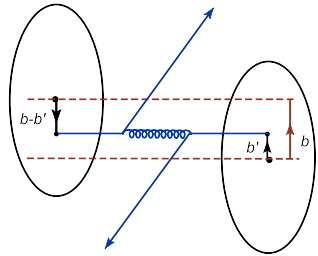}
\caption{Transverse overlap of the hadrons.}
\label{fig:myska_overlap}
\end{center}
\end{figure}

The evaluation of the overlap function $F(b)$ then depends
on the model of the internal hadron structure. Figure
\ref{fig:myska_overlap} explains the parameters used in the Eq. \ref{eq:myska_overlap},
where the overlap function is expressed as a convolution of the form
factor distributions of two incoming hadrons:

\begin{equation}
F(b) = \int d^2b' \rho_A(b') \rho_B(b-b').
\label{eq:myska_overlap}
\end{equation}

\noindent The overlap function also satisfies the normalization condition:
\begin{equation}
 \int \pi d^2b F(b) = 1.
\end{equation}

There are several possible models of hadron matter
offering the expressions for the function $\rho$, representing
the parton matter distributions in transverse space.
For example, the single Gaussian shape assumes the momentum
of individual partons to be distributed homogenously
in the whole hadron, while the double channel
Gaussian model predicts the hadron to contain a smaller
hard core surrounded by a shell made up of softer partons.

The question of the parton correlations inside the hadron is still
unanswered and can include only the spatial constraints, as in the
Poissonian model, or can contain all the possible correlations, such
as in momentum, spin, and color of the diparton system, see e.g. \cite{Mekhfi1}.

\section{Effective Cross Section}

The situation becomes more complex after involving the different
behaviors of every kind of parton in the theory. The overlap
functions need to carry two additional indices characterizing the
two interacting partons. The formula for the DPS cross section
(\ref{eq:myska_factorization}) has to be rewritten into the form:

\begin{equation}
 \sigma_D = \frac{m}{2} \sum_{ijkl} \int \sigma_S^{ij} \sigma_S^{kl}
F^i_k(b)F^j_l(b) d^2b,
\end{equation}

\noindent where $\sigma_S$ is the classical single parton scattering cross
section.

The integral over the impact parameter $b$ is then
marked as the geometrical coefficient $\Theta$

\begin{equation}
 \Theta_{kl}^{ij} = \int F^i_k(b)F^j_l(b) d^2b.
\end{equation}

\noindent These geometrical coefficients have to be taken into account
when one expects different partons to have a different
distribution in transverse space inside the hadron.

The simplest model ignores the differences in the geometrical
coefficients and approximates all the spatial
correlations as well as the uncertainty in the transverse
space parton distribution by the scale factor $\sigma_{eff}$ named
effective cross section:

\begin{equation}
 \sigma_{eff}^{-1} = \int \left[ F(b) \right]^2 d^2b.
\end{equation}

\noindent Eventually the DPS cross section \ref{eq:myska_factorization}
can be simplified to:

\begin{equation}
 \sigma_D = \frac{m}{2}\frac{\sigma_S^2}{\sigma_{eff}}.
\end{equation}

The prediction of the effective cross section value differs
according to the model of hadron structure. The first intuitive
estimation uses the average hadron radius $r \approx 0.6fm$. The
$\sigma_{eff}$ would be approximately $30mb$. The most promising
model currently is the double Gaussian model that predicts
$\sigma_{eff} \approx 11mb$ \cite{Treleani}. Table 1. summarizes the
effective cross section measurements by experiments at CERN and
Fermilab \cite{AFS}, \cite{UA2}, \cite{CDF93}, \cite{CDF97}.

\begin{table}[h!]
\caption{The effective cross section measurements.}
\begin{center}
\begin{tabular}{l|ccc} \hline
          & $\sqrt{s}$ & final state &  $\sigma_{eff} (mb)$ \\   \hline
AFS, 1986 & 63         & 4jets & $\approx 5$  \\
UA2, 1991 & 630        & 4jets & $\ge 8.3(95\%C.L.)$  \\
CDF, 1993 & 1800       & 4jets & $= 12.1^{+10.7}_{-5.4}$  \\
CDF, 1997 & 1800       & $\gamma+3jets$ & $= 14.5\pm1.7^{+1.7}_{-2.3}$  \\   \hline
\end{tabular}
\end{center}
\end{table}

\section{Double Drell-Yan Process}

\begin{figure}[h!]
\begin{center}
\includegraphics[width=4cm]{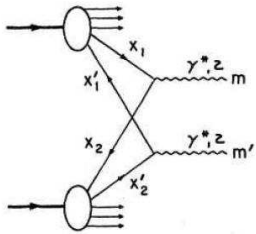}
\caption{The DPS mechanism of $\gamma^*/Z$ pair production.}
\label{fig:myska_DDY}
\end{center}
\end{figure}

The possibility of creation of a neutral vector boson pair via the
double parton scattering mechanism has already been suggested many
years ago \cite{Halzen}, \cite{Mekhfi2}. In this process, the two
independent s-channel quark-antiquark annihilations occur in one
hadron-hadron interaction, see Fig. \ref{fig:myska_DDY}

On the basis of the previous theoretical description, the double
Drell-Yan production cross section is predicted to be approximately
$\sigma_{DDY} =
\frac{\sigma_S^2}{2\sigma_{eff}}\approx\sigma_S.10^{-7}\approx0.1fb$,
where only the electron decay channel is taken into account. The
single parton production cross section was taken to be $\sigma_S
\approx 1.7nb$.

The motivation for this study was the promise of the cleanest signal
in the detector \cite{Dress} in comparison to the multi-jet
processes, even at the cost of a small cross section.

The source of the data for this preliminary analysis was the
Herwig++ 2.3.0 Monte Carlo event generator \cite{Herwig}. The model
of MPS is already implemented in Herwig++. The studied process
consisted of the primary hard subprocess of type $MEqq2gZ2ll$ and of
the additional hard subprocess of the same type with the
multiplicity equal to one. The full matrix element was used. The
final state was created by the two electron-positron pairs. The MPS
model of Herwig++ performs the calculation of the number of
sub-processes independently and adds the general $QCD2\rightarrow2$
matrix element to them.

\begin{figure}[h!]
\begin{center}
\includegraphics[width=4cm]{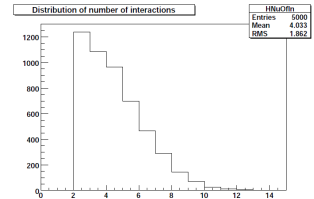}
\caption{Distribution of total number of parton interactions for
5000 proton-proton events at $\sqrt{s} = 10 TeV$. There are always
at least two $q\bar{q}\rightarrow \gamma^*/Z \rightarrow e^+e^-$
interactions.} \label{fig:myska_interactions}
\end{center}
\end{figure}

The simulation was performed for proton-proton collisions at
$\sqrt{s} = 10 TeV$. The distribution of the total number of parton
interactions for 5000 proton-proton events is shown in Fig.
\ref{fig:myska_interactions} The shape of the distribution
corresponds to the Poissonian distribution with a little longer
tail. The average number of interactions has been found to be four.

Other kinematical distributions seem to be similar to the single
parton but are not investigated in detail in this stage of the
study. There are few suspicions that the Monte Carlo implementation
of the MPS model still needs to be tuned. The production cross
section is not provided because the MPS is treated as the underlying
event activity and can not be set as part of the production cross
section calculation.

\section{Summary and Conclusions}

Multiple parton scattering was experimentally measured by several
collaborations and its existence was proven. The previous
experiments aimed their attention at the multi-jet final states or
at photon-plus-jets final state, where the production cross section
is sufficiently large and the MPS processes contribute significantly
especially at a low $p_T$ region. The motivation to search for the
other processes, like vector boson pair production, is increased by
the approaching LHC era.

The theoretical model describing MPS is basically developed, but it
suffers from many sources of uncertainty. Especially the internal
parton correlations, which is one of the unknown effects that needs
to be investigated further. MPS is also a unique tool for
geometrical coefficient measurement. It may resolve the size of
hard/soft component of the proton, and may bring about new many-body
parton distribution functions.

The preliminary predictions for the double Drell-Yan process shows a
very low cross section for the lepton decay channel and it will not
be measurable at LHC. For the same reason, it will not be a source
of background for Higgs or for di-boson production.

\end{document}